\documentclass[aps,prl,twocolumn,showpacs,superscriptaddress]{revtex4}
\usepackage{graphicx}

\usepackage{bm}
\usepackage{amsmath}

\newcommand{\dto}{Dy$_{2}$Ti$_{2}$O$_{7}$}

\newcommand{\dytio}{Dy$_{2}$Ti$_{2}$O$_{7}$}

\begin{document}


\title{Kagom\'{e} ice state in the dipolar spin ice Dy$_{2}$Ti$_{2}$O$_{7}$ }



\author{Y.~Tabata}
\affiliation{Graduate School of Science, Osaka University, Toyonaka, Osaka 560-0043, Japan}

\author{H.~Kadowaki}
\affiliation{Department of Physics, Tokyo Metropolitan University, Hachioji-shi, Tokyo 192-0397, Japan}

\author{K.~Matsuhira}
\affiliation{Department of Electronics, Kyushu Institute of Technology, 
Kitakyushu 804-8550, Japan}

\author{Z.~Hiroi}
\affiliation{Institute for Solid State Physics, University of Tokyo, Kashiwa, Chiba 277-8581, Japan}

\author{N.~Aso}
\affiliation{NSL, Institute for Solid State Physics, University of Tokyo, Tokai, Ibaraki 319-1106, Japan}

\author{E.~Ressouche}
\author{B.~F{\aa}k}
\affiliation{CEA, D\'{e}partement de Recherche Fondamentale sur la Mati\'{e}re Condens\'{e}e, SPSMS, 38054 Grenoble, France}


\date{\today}

\begin{abstract}
We have investigated the kagom\'{e} ice behavior of 
the dipolar spin-ice compound {\dytio} in 
magnetic field along a [111] direction using neutron scattering 
and Monte Carlo simulations. 
The spin correlations show that the kagom\'{e} ice 
behavior predicted for the nearest-neighbor (NN) interacting 
model, where the field induces dimensional reduction 
and spins are frustrated in each two-dimensional kagom\'{e} lattice, 
occurs in the dipole interacting system. 
The spins freeze 
at low temperatures within the 
macroscopically degenerate ground states of the 
NN model. 
\end{abstract}

\pacs{75.10.Hk, 05.50.+q, 75.25.+z}

\maketitle

Geometrically frustrated spin systems have been investigated for decades 
because of their fascinating magnetic properties 
\cite{Bramwell01,Gardner_TTO,Gaulin_TMO,Mirebeau_TTO,Lee_ZCO}. 
An Ising model on a cubic pyrochlore lattice consisting of 
spins parallel to local $\langle 111 \rangle$ 
easy axes was shown to exhibit macroscopic degeneracy 
in the ground states \cite{Bramwell01,Harris_HTO,Ramirez_DTO}
in the same way as the proton disorder 
in water ice does \cite{Pauling1935}. 
The local easy axis, \textit{ferromagnetic} nearest-neighbor 
(NN) exchange interaction, and the geometry of corner sharing 
tetrahedra give rise to frustration in this system. 
Since the local ``two-in and two-out'' spin configuration of the ground 
states simulates the ``ice rule'' in water ice, 
the Ising model is referred to as ``spin ice'' 
\cite{Bramwell01}. 

Frustrated behavior attributable to the 
NN ferromagnetic spin-ice model was discovered 
in pyrochlore oxides, 
e.g. R$_2$Ti$_2$O$_7$, R$_2$Sn$_2$O$_7$ (R=Dy, Ho)
\cite{Harris_HTO,Ramirez_DTO,kadowaki02,matsuhira00}, 
where the observed residual entropy agrees 
with the Pauling estimate for water ice \cite{Ramirez_DTO}. 
However, spins in these oxides interact 
dominantly by the dipolar interaction. 
Hence, they are represented by another Ising model 
with a dominant dipolar interaction, 
referred to as dipolar spin ice \cite{Hertog_PRL}. 
It was a puzzle why the 
dipolar spin ice shows the 
spin ice behavior predicted for the NN model. 
This was solved by 
Monte Carlo (MC) simulation studies 
\cite{Hertog_PRL,Melko_PRL,Melko_JPCM}, 
showing that the dipolar interaction brings about an 
effective NN ferromagnetic coupling. 
In addition, it was shown that the long-range 
dipolar interaction only slightly 
lifts the degeneracy of the ground state manifold 
of the NN spin-ice model, 
and spins freeze within this manifold 
at low temperatures and never reach the true ground 
state \cite{Melko_PRL,Melko_JPCM}. 
The simulations also reproduce experimental observations 
of the dipolar spin-ice compounds 
\cite{Hertog_PRL,Bramwell_hto,Melko_JPCM}. 
More recently, an elegant analytical understanding 
has been exploited using an argument of projective equivalence 
\cite{Isakov_mode}.

\begin{figure}
\begin{center}
\includegraphics[width=7cm,clip]{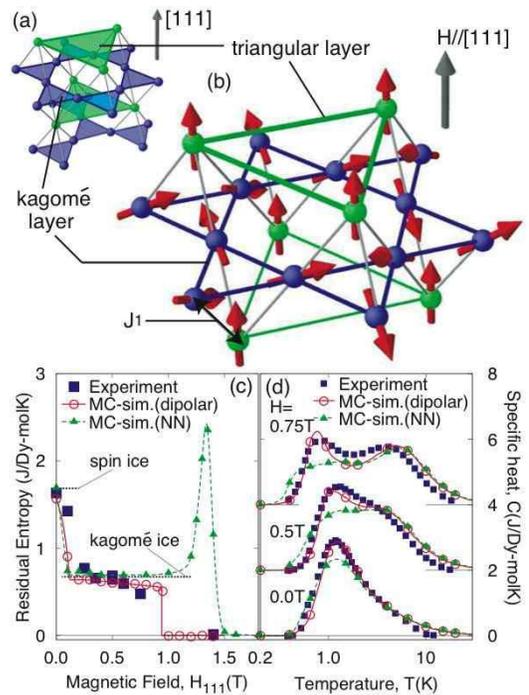}
\end{center}
\caption{(color online) 
(a) The pyrochlore lattice is alternating stacking of 
kagom\'{e} and triangular layers along a [111] direction. 
(b) A spin configuration of the kagom\'{e} ice state is shown. 
(c) The observed \cite{Hiroi_resS} and calculated residual entropies 
are plotted as a function of magnetic field. 
The calculated entropy for the NN model \cite{Isakov_kagome} 
is $S(T=0.1 \text{K})$, which is temperature dependent 
in the peaked region. 
Dotted lines represent the precise values of the spin ice and 
kagom\'{e} ice states for the NN model 
\cite{Moessner_kagome,Udagawa_kagome}.
(d) Temperature dependence of observed \cite{Hiroi_resS} and 
calculated specific heat at $H = 0, 0.5, 0.75$ T.
\label{Macros}}
\end{figure}

By applying magnetic fields to the NN spin-ice model 
the macroscopic degeneracy of the ground states can be 
partly or fully lifted depending on the field direction 
and magnitude \cite{spin-ice_field,Ramirez_DTO}. 
For a field parallel to a [111] direction, along which the 
pyrochlore lattice is alternating stacking of 
kagom\'{e} and triangular layers [see Fig.~\ref{Macros}(a)], 
the field with intermediate strengths partly lifts the degeneracy and 
simultaneously induces decoupling of spins between the layers.
In these ground states, spins on the triangular layers are parallel to 
the field, and spins on the kagom\'{e} layers retain 
macroscopic degeneracy preserving the ``two-in and two-out'' 
configuration at each tetrahedron [see Fig.~\ref{Macros}(b)] 
\cite{Matsuhira_kagome}. 
The ground states on each kagom\'{e} layer can be mapped 
onto those of an exactly solvable dimer model in two dimensions 
\cite{Moessner_kagome,Udagawa_kagome,Isakov_kagome}. 
At low temperatures, by neglecting excited states with an energy gap, 
one can expect ice-like behavior of the spins on the kagom\'{e} lattice, 
referred to as ``kagom\'{e} ice" \cite{Matsuhira_kagome}. 

For the dipolar spin ice {\dto}, experiments 
with field along [111] were performed 
\cite{Matsuhira_kagome,Hiroi_resS,Higashinaka_kagome,Sakakibara_M}. 
It was shown that the magnetization plateau is consistent with the 
predictions of the kagom\'{e} ice behavior for the 
NN model \cite{Matsuhira_kagome,Sakakibara_M}. 
However, the residual entropy plateau, another important 
point for proving the kagom\'{e} ice, has not been definitely 
shown by the specific heat measurements 
\cite{Matsuhira_kagome,Hiroi_resS,Higashinaka_kagome} 
because of experimental difficulties \cite{Hiroi_resS}. 
In addition, effects of the long range nature of 
the dipolar interaction have not been 
clarified for the kagom\'{e} ice state. 
In this work, we performed neutron scattering experiments 
on {\dto} with field along [111] in the 
conjectured kagom\'{e} ice regime, aiming at 
microscopic demonstration of the  kagom\'{e} ice state.
We compared the neutron data with MC simulations 
based on the dipolar spin-ice model 
\cite{Hertog_PRL,Melko_JPCM}. 
The simulations reproduce the main characteristics of the kagom\'{e} ice:
the residual entropy, the spin correlations 
close to the wave vector $(\frac{2}{3},-\frac{2}{3},0)$, 
and the field-induced dimensional reduction, being consistent with 
observations. 

Most of neutron scattering experiments were performed on the 
triple-axis spectrometer GPTAS installed at JRR3, 
Japan Atomic Energy Agency. 
A single crystal sample of {\dto} used in this work was 
$22 \times 3.1 \times 0.58$ mm$^3$ in size. 
Its long direction is parallel to a [111] direction, 
which ensures negligibly small demagnetization effect. 
The sample was mounted in a dilution refrigerator 
so as to measure the scattering plane perpendicular to the 
[111] direction. 
A scan along the [111] direction was carried out 
on the lifting-arm two-axis diffractometer 
D23 CEA-CRG at Institute Laue-Langevin.
All the data shown are corrected for background and absorption. 
The MC simulations were carried out using a supercomputer 
at Institute for Solid State Physics, University of Tokyo. 
A standard Metropolis algorithm with single-spin-flip dynamics 
was used in our simulations. 
The long-range dipolar interaction was handled by a method 
\cite{Kadowaki_MC} 
equivalent to the Ewald method \cite{Hertog_PRL,Melko_JPCM}. 
We carried out the simulations up to 38400 spins 
with up to $6 \times 10^{5}$ MC steps per spin.

We adopted the Hamiltonian used 
in Refs.~\cite{Hertog_PRL,Ruff_inH} 
\begin{align}
  \mathcal{H} = & -\mu _{\text{eff}} \sum _{i,a} {\bm S}_{i}^{a} \cdot {\bm H} - \sum _{\langle (i,a),(j,b) \rangle} J_{i,a;j,b}{\bm S}_{i}^{a} \cdot {\bm S}_{j}^{b} \notag
  \\
   & +Dr_{\text{nn}}^{3} 
\sum \left[
\frac{{\bm S}_{i}^{a} \cdot {\bm S}_{j}^{b}}
{\mid {\bm R}_{ij}^{ab} \mid^{3}} 
- \frac{3({\bm S}_{i}^{a} \cdot {\bm R}_{ij}^{ab})({\bm S}_{j}^{b} \cdot {\bm R}_{ij}^{ab})}{\mid {\bm R}_{ij}^{ab} \mid^{5}} \right], \notag
\end{align}
where ${\bm S}_{i}^{a}$ represents the spin vector 
parallel to the local $\langle 111 \rangle$ easy axis 
with $|{\bm S}_{i}^{a}|=1$ at the 
sublattice site $a$ in the unit cell 
of the fcc lattice site $i$. 
The first term is the Zeeman interaction between 
the spins possessing the effective moment 
$\mu_{\text{eff}} \simeq 10$ $\mu_{\text{B}}$ 
and the magnetic field ${\bm H}$. 
The second and third terms are the exchange and dipolar 
interactions, respectively.
For the NN ferromagnetic spin-ice model 
there is only one exchange coupling 
constant $J_{1}>0$ and $D=0$.

\begin{figure*}
\includegraphics[width=16cm,clip]{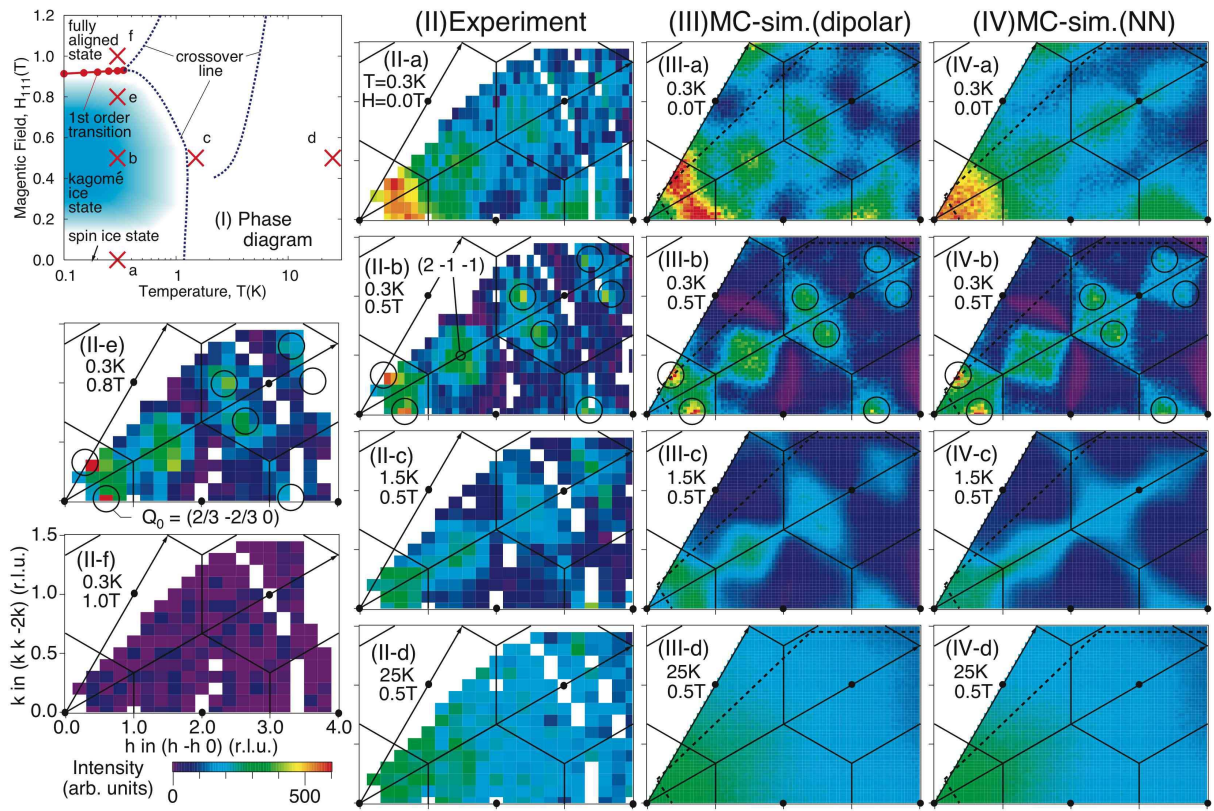}
\caption{(color) 
(I) $HT$ phase diagram of {\dto} obtained from the 
magnetization and the specific heat measurements 
\cite{Sakakibara_M,Hiroi_resS}. 
The kagom\'{e} ice state is observed in the blue region. 
The red solid line represents the first-order spin-flop transition. 
(II, III, IV) Neutron intensity patterns in the scattering plane perpendicular to the 
[111] direction are shown for several temperatures and fields 
denoted by crosses in (I). 
In (II-a)$\sim$(II-f) experimental results for {\dto} are shown. 
Calculations using the MC simulation based on the dipolar and 
NN spin ice models are shown in (III-a)$\sim$(III-d) and (IV-a)$\sim$(IV-d), 
respectively. 
\label{map}}
\end{figure*}

It is well established that the low temperature 
behavior of {\dto} can be approximately reproduced by a 
single-spin-flip MC simulation based on the dipolar 
spin-ice model 
\cite{Hertog_PRL,Melko_PRL,Melko_JPCM,Ruff_inH} 
except a few observations. 
For this model the dipolar interaction with $D = 1.41$ K 
and the Zeeman interaction are the dominant terms. 
The weaker exchange interactions consist of the NN 
antiferromagnetic interaction with $J_{1} = -3.72$ K, 
and smaller second and third neighbor couplings with 
$J_{2}= 0.1$ and $J_{3}= -0.03$ K, respectively. 
The very weak $J_{3}$ becomes important for determining the ground state 
within the degenerate states when the field is along [112], because larger interactions 
balance each other \cite{Ruff_inH,Higashinaka_112}.
Another balancing occurs for the [111] field close 
to the spin flop transition [see Fig.~\ref{map}(I)], 
which allowed us to determine $J_{2}$ 
by a fit of the $T$ dependence of the specific heat 
$C(H = 0.75 \text{T})$ [see Fig.~\ref{Macros}(d)] to the simulation
\cite{Kadowaki_MC}. 

The kagom\'{e} ice behavior in {\dto} was suggested by the 
residual entropy measurement \cite{Matsuhira_kagome,Hiroi_resS} 
for fields close to $H=0.5$ T as shown in Fig.~\ref{Macros}(c). 
We calculated the residual entropy using 
MC simulation of the dipolar spin-ice model, where 
$C(H)/T$ is integrated over $0.2 \leq T \leq 200$ K. 
The MC calculations are shown in 
Fig.~\ref{Macros}(c) and (d) together with 
those using the NN spin-ice model with ferromagnetic 
$J_{1} = 4.5$ K \cite{Isakov_kagome}. 
The computation agrees with the experiment 
around $H \sim 0.5$ T, and shows more clearly the plateau 
behavior than the experiment \cite{Hiroi_resS}. 
The low temperature entropy $S(T=0.1 \text{K})$ of the NN spin-ice model 
\cite{Isakov_kagome}, showing a peaked structure at 
a higher field ($H\sim 1.35$ T), indicates that the discontinuous 
decrease of the residual entropy at $H \sim 0.9$ T, 
accompanying the spin flop transition \cite{Sakakibara_M}, 
is brought about by the long-range dipolar interaction. 

The plateau of the residual entropy around 
$H \sim 0.5$ T of the dipolar spin-ice model 
is slightly smaller than that of the NN model. 
This suggests that the kagom\'{e}-ice ground-state 
manifold of the NN model is weakly lifted 
by the dipolar interaction. 
For the origin of the kagom\'{e} ice behavior of {\dto} 
one naturally expects that spins freeze 
within these nearly degenerate ground states 
owing to the high energy barriers for breaking 
the ``two-in and two-out'' ice rule \cite{Melko_PRL}. 
This mechanism can be microscopically investigated 
by observing spin correlations 
using neutron scattering techniques. 

In Fig.~\ref{map}(II) we show neutron intensity patterns in 
the scattering plane perpendicular to the [111] direction 
at several points in the $HT$ plane denoted by crosses in 
Fig.~\ref{map}(I). 
The low temperature data at $T=0.3$ K show 
definite differences between the spin ice state at $H=0$ [(II-a)], 
the kagom\'{e} ice state at $H=0.5, 0.8$ T [(II-b),(II-e)], 
and the fully aligned state at $H=1$ T [(II-f)], where 
very weak scattering intensity implies that very small 
spin fluctuations are left above the spin flop transition. 
We note that the intensity peak 
observed for the kagom\'{e} ice state [(II-b),(II-e)] 
at the wave vector $\bm{Q}_0= (\frac{2}{3},-\frac{2}{3},0)$, 
denoted by black circles, is a characteristic of the 
kagom\'{e} ice state. 
This spin correlation was found in the exact ground states 
of the NN spin-ice model \cite{Moessner_kagome}. 

In Fig.~\ref{map}(III) and (IV) we show scattering intensities 
calculated by MC simulations based on the dipolar and 
NN spin-ice models, respectively. 
The simulations for the spin ice state [(III-a),(IV-a)], 
which are consistent with previous results 
\cite{Bramwell_hto,Fennel_dto}, 
confirm that the agreement with experiments [(II-a)] is improved 
by taking account of the dipolar interaction. 
In contrast to this, the simulations 
for the kagom\'{e} ice state [(III-b),(IV-b)] show 
almost the same pattern to the naked eye, and agree well 
with the experiment [(II-b)]. 
As noticed previously, the field induces decoupling of the 
spins between the kagom\'{e} layers for the 
NN model. 
This field-induced dimensional-reduction is confirmed 
by no $Q$ dependence of the calculated intensity pattern [(IV-b)] 
along the [111] direction. 
For the dipolar model, we also observed little 
$Q$ dependence of the calculated intensity pattern [(III-b)] 
along the [111] direction, and hence conclude that 
the spin correlations show quasi-two dimensional 
character in the kagom\'{e} ice state. 
As temperature is increased at the typical field $H=0.5$ T of 
the kagom\'{e} ice state, the experimental intensity patterns [(II-c),(II-d)] 
are consistent with the simulations [(III-c,d),(IV-c,d)]. 
From these results 
we conclude that the degeneracy of the kagom\'{e} ice 
ground states of the NN spin-ice model
is almost identically preserved for the dipolar spin-ice model 
and for {\dytio}, providing microscopic evidence for the 
kagom\'{e} ice behavior of this compound. 
\begin{figure}
\includegraphics[width=8cm,clip]{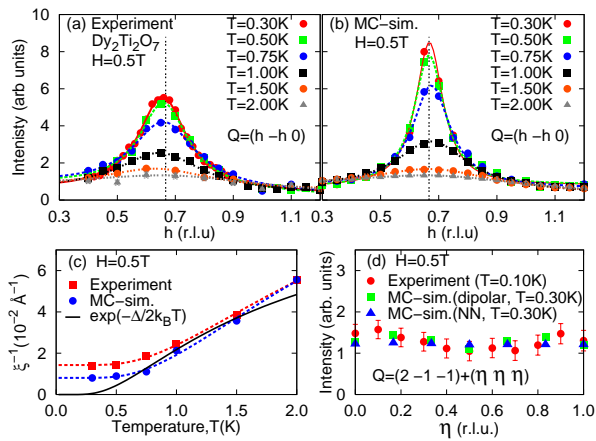}
\caption{(color online) 
$Q$-scans along the $(h,-h,0)$ line through ${\bm Q}_0$ at $H= 0.5$ T: 
(a) experimental and (b) 
MC simulation results based on the dipolar model. 
Curves are fits to Lorentzian functions. 
(c) Temperature dependence of the inverse correlation length $\xi^{-1}$
obtained by the fits. 
The dotted lines are guides to eyes. 
The solid line represents the exponential $T$ dependence 
$\exp (-\Delta/2k_{B}T)$ described in the text.
(d) $Q$-scan along the [111] direction from $(2\bar{1}\bar{1})$ 
[Fig.~\ref{map}(II-b)].
\label{cor}}
\end{figure}

In order to study the spin correlations around ${\bm Q}_0$ 
at $H=0.5$ T, we performed scans along the $(h,-h,0)$ line 
and plot the results in Fig.~\ref{cor}, where 
the MC calculations of the dipolar model are also shown. 
One can see from Fig.~\ref{cor}(a) that 
the spin correlations of {\dytio} cease to develop around 
$T=0.5$ K. 
This disagrees with the theory of the 
NN model \cite{Moessner_kagome}, where 
it was argued that the correlation 
length diverges exponentially as $\exp (\Delta/2k_{B}T)$ 
($T \rightarrow 0$), 
where $\Delta$ is a gap energy, in the thermal equilibrium state. 
The non divergence of the spin correlations for {\dytio} can be 
ascribed to spin freezing due to the barrier for 
breaking the ice rule. 
As discussed in Ref.~\cite{Melko_PRL} for the spin ice state, 
the single-spin-flip dynamics of the MC simulation 
become non-ergodic at low temperatures, 
and hence, the simulated system does not reach 
a thermal equilibrium state within practical MC steps, 
as well as {\dytio} does not within experimental time 
scales. 
In fact, the simulated system also show freezing around $T=0.7$ K, 
which does not necessarily agree with the experiment 
because of difference of the spin-flip dynamics 
between the simulated system and {\dytio}. 
To observe the two-dimensional spin-correlations 
in the kagom\'{e} ice state, 
we performed a $Q$-scan along the [111] direction 
from $(2\bar{1}\bar{1})$ [see Fig.~\ref{map}(II-b)]. 
The result together with the MC simulations of the two models 
are shown in Fig.~\ref{cor}(d). 
These experimental data, being consistent with 
the simulations within the experimental error, 
suggest the two dimensional nature of 
the kagom\'{e} ice state in {\dytio}.

Finally we note that 
a surprising fact revealed by the present numerical 
study is the high similarity of the spin correlations, 
shown in Fig.~\ref{map}(III-b)$\sim$(III-d) and (IV-b)$\sim$(IV-d), 
between the NN and dipolar models 
especially for the kagom\'{e} ice state. 
In view of the theoretical work \cite{Isakov_mode,Hertog_PRL} 
explaining the equivalence of the NN and dipolar models 
for the spin ice state, there may be a similar 
enlightening argument also for the kagom\'{e} ice state. 

In summary, we have investigated the kagom\'{e} ice behavior 
and field-induced dimensional reduction 
of the dipolar spin ice {\dto} in magnetic field along 
the [111] direction using neutron scattering. 
The observed spin correlations were analyzed by 
MC simulations based on the dipolar and NN spin-ice models. 
For fields around $H \sim 0.5$ T 
spins freeze within 
the degenerate kagom\'{e}-ice ground states of the NN model, 
which are weakly lifted by the dipolar interaction, 
and consequently exhibit kagom\'{e} ice behavior. 



\end{document}